\documentclass{PoS}

\title{The Infrared behavior of SU(3) Nf=12 gauge theory -about the existence of conformal fixed point- }
\ShortTitle{The IR behavior of SU(3) Nf=12 gauge theory -about the existence of conformal FP-}

\author{\speaker{Kenji~Ogawa}$^a$\thanks{E-mail: ogawak@mail.nctu.edu.tw},
Tatsumi~Aoyama$^b$, Hiroaki~Ikeda$^c$,Etsuko~Itou$^d$, Masafumi~Kurachi$^{b}$, C.-J.~David~Lin$^{ae}$, Hideo~Matsufuru$^f$, Hiroshi~Ohki$^{b}$, Tetsuya~Onogi$^d$, Eigo~Shintani$^g$ and Takeshi~Yamazaki$^b$\\
$(a)$Institute of Physics, National Chiao-Tung University, Hsinchu 300, Taiwan\\
$(b)$Kobayashi-Maskawa Institute for the Origin of Particles and the Universe, Nagoya University, Nagoya, Aichi 464-8602, Japan\\
$(c)$School of High Energy Accelerator Science, The Graduate University for Advanced Studies (Sokendai), Tsukuba 305-0801, Japan\\
$(d)$Department of Physics, Osaka University, Toyonaka 560-0043, Japan \\
$(e)$Division of Physics, National Center for Theoretical Sciences, Hsinchu 300, Taiwan\\
$(f)$High Energy Accelerator Research Organization (KEK), Tsukuba 305-0801, Japan\\
$(g)$RIKEN-BNL Research Center, Brookhaven National Laboratory, Upton, NY 11973, USA\\ 
}

\usepackage{graphicx}
\usepackage{amsmath, amssymb}
\usepackage{amssymb}

\newcommand{\beq}{\begin{equation}}
\newcommand{\eeq}{\end{equation}}
\newcommand{\bea}{\begin{eqnarray}}
\newcommand{\eea}{\end{eqnarray}}
\newcommand{\bev}{\begin{verbatim}}
\newcommand{\eev}{\end{verbatim}}

\newcommand{\BAN}{\begin{eqnarray*}}
\newcommand{\EAN}{\end{eqnarray*}}

\abstract{
Incorporated with twisted boundary condition, Polyakov loop correlators can give a definition of the renormalized coupling.
We employ this scheme for the step scaling method (with step size s = 2 ) in the search of conformal fixed point of SU(3) gauge theory with 12 massless flavors.
Staggered fermion and plaquette gauge action are used in the lattice simulation with six different lattice sizes, L/a = 20, 16, 12, 10, 8 and 6.
For the largest lattice size, L/a = 20, we used a large number of Graphics Processing Units (GPUs) and accumulated 3,000,000 trajectories in total.
We found that the step scaling function $\sigma (u)$ is consistent with u in the low-energy region.
This means the existence of conformal fixed point.
Some details of our analysis and simulations will also be presented.
}

\FullConference{ The XXIX International Symposium on Lattice Field Theory - Lattice 2011\\
July 10-16, 2011\\
Squaw Valley, Lake Tahoe, California}

\begin{document}

\section{Introduction}
Gauge theories with many flavors are paid attention,
as candidates of beyond the standard model physics
which might give a solution to the hierarchy problem.
In this work, we investigate a model which consists of SU(3) gauge theory coupled to 12 massless fermion flavors in the fundamental representation.
We study the running of coupling constant in this system.
Our main concern here is the existence of the infrared fixed point, which is a scheme-independent property of the model.
In the infrared properties of such a system,
nonpertubative effects are dominant,
and the numerical simulations with the lattice regularization are very useful tools, as demonstrated in QCD.
So far, this model has been studied by several groups using lattice simulations with different methods
\cite{Appelquist:2009ty}-\cite{Hasenfratz:2010fi}.
At this moment, they have not yet reached a clear consensus.
In this work, we adopt the Twisted Polyakov Loop scheme\,\cite{lw:1986}\cite{deDivitiis:1993hj}\cite{deDivitiis:1994yp} to study the running coupling constant. This method is different from those used by others. 
We show result which supports the existence of conformal fixed point.
%
%

We already published several articles
using this method \cite{Aoyama:2011ry}-\cite{Ohki:2010sr}.
This work is the continuation of Refs.\,\cite{Bilgici:2009nm}\cite{Itou:2010we}.
The data set used here is almost the same as our work reported in Ref.\,\cite{Aoyama:2011ry}.

This paper is organized as follows.
In Sec.\,\ref{sec:twbc}, we give a review on twisted boundary conditions and the Twisted Polyakov Loop scheme,
which we used to calculate the renormalized coupling constant.
In Sec.\,\ref{sec:setup}, we explain the details of the set up of our simulations.
The results are given in Sec.\,\ref{sec:results}.
Summary and comments are given in Sec.\,\ref{sec:summary}.
\section{Twisted Polyakov Loop Scheme}\label{sec:twbc}
In this section, we explain the twisted boundary condition and 
our definition of renormalized coupling constant using Polyakov loops.
The twisted boundary condition results in  infrared cut-offs in both gluon and fermion propagators.
Then, the calculation of step scaling function is feasible without using the Schroedinger Functional method,
in which Dirichlet boundary conditions are required.

The twisted boundary condition for the link variables is given as
\beq
U_\mu( x + \hat \nu L_\nu/a )
= \Omega_\nu U_\mu( x ) \Omega_\nu^\dagger.
\eeq
We apply this condition only for $ \nu = 1, 2 $ direction.
Here, $\Omega_i (i=1,2) $ are SU(3) matrices which satisfy the conditions
$
\Omega_1 \Omega_2 
= e^{i2\pi/3} \Omega_2 \Omega_1$
,
$
\Omega_\mu \Omega_\mu^\dagger = 1
$
,
$
\left ( \Omega_\mu \right )^3 = 1  
$
and
$
\mathrm{Tr} 
\left [ \Omega_\mu \right ] = 0.
$
Their explicit form can be 
$[\Omega_1]_{ab} = \delta_{a,\!\!\!\mod\!\!(b+2,3)+1}$ and 
$\Omega_2 = \mathrm{diag}(e^{-i 2 \pi /3} , e^{ i 2 \pi /3} , 1 )$.
For $\nu = 3, 4$ direction, we implement the periodic boundary condition, $\Omega_3 = \Omega_4 = {\bf 1}$.
%
%
%

The Polyakov loops in the twisted directions are
\beq
P_1 ( x_2 , x_3 , x_4 )
=
\mathrm{Tr}
\left ( 
\left [
\prod_j
U_1
\left (
x_1 = j , x_2 , x_3 , x_4
\right )
\right ]
\Omega_1 e^{i 2 \pi x_2 a / ( 3 L_2 ) }
\right ).
\eeq
The extra terms are added
to maintain gauge invariance and  translation invariance.
The coupling constant can be defined as the ratio of the correlators of Polyakov loops in the twisted direction and periodic direction,
\beq
\frac{ 
\langle 
P_1( x_4 = 0 )^\dagger P_1( x_4 = L_4/(2 a) ) 
\rangle
}{ 
\langle 
P_3( x_4 = 0 )^\dagger P_3( x_4 = L_4/(2 a))
\rangle
} = k \bar g^2,
\eeq
with the factor $k = 0.03184\dots $ which is given by calculating one-gluon-exchange diagram.
In the actual analysis, 
we average over data that are related by translation and 90-degree rotation, to reduce the statistical errors.
To apply the twisted boundary condition to the system with fermions,
we need to introduce another degree of freedom, "smell", 
to ensure the single-valuedness of $\psi\,(~x~+~L_1\,\hat 1~+~L_2\,\hat 2~)$ under the application of two boundary twisting, in $x_1$ and $x_2$ directions, with the different ordering.
The boundary condition is given by,
\beq
\psi^a_\alpha( x + L_\nu \hat \nu ) = e^{i\pi/3} \Omega_\nu^{ab} \psi^b_\beta \left ( \Omega_\nu \right )^\dagger_{\beta \alpha}.
\eeq
The indices $\alpha$, $\beta$ are labels for the "smell" degree of freedom.
The factor $e^{i\pi/3}$ are multiplied to eliminate the zero-momentum mode in $\nu= 1,2$ direction.
For $\nu = 3, 4$ direction, we implement the periodic boundary condition, $\psi( x+ L_\nu \hat \nu ) = \psi( x )$. 
The number of "smell"s is the same as the number of colors. 
The "smell" index does not combine with the gauge field, and it can be considered as the another flavor index.
Under this condition,
the number of flavors should be multiples of $4({\rm taste}) \times 3({\rm smell})$
for the dynamical simulations with staggered fermions 
without taking roots of fermion determinant.
\section{Setting of Simulations}\label{sec:setup}
To study the scale dependence of the coupling constant by lattice simulations, 
we measure the step scaling function, which is defined as
\beq
\sigma(u,s) = \left . {\bar g}^2 \left ( \frac{1}{ s L } \right ) \right |_{ u = {\bar g}^2 \left ( \frac{1}{ L } \right ) }.
\label{eq:ssf}
\eeq
This is calculated by taking the continuum limit 
of the lattice step scaling function $\Sigma( u , s , L/a )$.
\beq
\Sigma( u , s , L/a ) = 
\left . \bar g^2_\mathrm{lattice} \left ( \beta , sL/a \right ) \right |_{ u = \bar g^2_\mathrm{lattice} \left ( \beta  , L/a  \right ) },
\label{eq:ssf_lat}
\eeq
\beq
\sigma( u , s ) = \lim_{a/L\rightarrow 0} \Sigma( u , s , L/a )~~~{\rm with~}L{\rm ~fixed}.
\eeq
In this work, we fix $s = 2$, and define $\sigma ( u ) = \sigma( u , 2 )$ and $\Sigma( u , L/a ) = \Sigma( u , 2 , L/a )$.

We have done the hybrid Monte Carlo (HMC) simulation for SU(3) gauge theory with 12 dynamical flavors by using staggered fermion.
Twisted boundary conditions are applied for the gauge field and fermion field as explained in the previous section.
For the gluon action, we use plaquette gauge action at $\beta = 4.5 \sim 100$.
In all simulations, we set the mass of fermions to be zero.
We used hyper-cubic box $L_1/a = L_2/a = L_3/a = L_4/a = L/a$
with the lattice size $L/a$ between 6 and 20.
Then the lattice step scaling function is calculated by using the combinations of $(L/a,sL/a) = $(6,12), (8,16) and (10,20).
The simulation was done by selecting the starting configuration to be in the true vacuum, 
which leads to a non-zero phase of Polyakov loop ( $\arctan( \mathrm{Im}(P_\mu) /\mathrm{Re}(P_\mu) ) \sim \pm 2\pi/3$ ) in $\mu = 3, 4$ direction, as discussed in Ref.\,\cite{Itou:2010we}.
The simulation parameters are summarized in Table\,\ref{table:numtraj}.

For the largest lattice $ L/a = 20 $, we perform the HMC simulations on GPUs.  
The simulation code was developed by using CUDA \cite{cuda}. 
We used around $50 \sim 100$ GPUs at the same time.
Most of them are Tesla 1060.
We used one GPU for one Markov chain.
For each parameter set, we produced 10 to 50 Markov chains at the same time, starting from the same seed configuration which is thermalized for that parameter set. To remove the correlation between the different Markov chains with the same starting configuration, we discarded 500 trajectories in the beginning of each Markov chain. The performance of the simulation with GPU is around 25 GFlops on average (sustained). 

To extract $\sigma(u,s)$\ in Eq\,(\ref{eq:ssf}), in practice,
we have done simulations with all beta values listed in Table\,\ref{table:numtraj} first, then we perform the interpolation in $\beta$ for the renormalized coupling at fixed $L/a$.
The lattice step scaling functions\,(\ref{eq:ssf_lat}) are calculated from the interpolated values of the renormalized coupling constant.
For this interpolation, we used non-decreasing polynomials with degree $(2 n + 1)$, 
\beq
f(x) = \int dx \left ( \sum_{l=0}^{n} \, c_l \, x^l \right )^2
= h_0 + h_1 x + \cdots + h_{2n+1} x^{2n+1},
\label{eq:fitfunc}
\eeq
with $x = 1 / \beta$. Here, we choose $h_0$ to be $0$ and $h_1( = c_0^2)$ to be $6$ to match with the perturbative expansion. Thus the number of free parameters becomes $ (n - 1) $.
  
\begin{table}[]
\center
\begin{tabular}{|c|c|} \hline
Lattice Size & $\beta$ \\ \hline
$6^4$  &  4.5,  4.7,  5.0,  5.5,  6.0,  6.5,  7.0,  8.0,  9.0,  10.0,  12.0,  14.0,  16.0,  18.0,  20.0 , 50.0, 100.0  \\ \hline 
$8^4$  &  4.5,  4.7,  5.0,  5.5,  6.0,  6.5,  7.0,  8.0,  9.0,  10.0,  12.0,  14.0,  16.0,  18.0,  20.0, 50.0, 99.0   \\ \hline 
$10^4$ &  4.5,  5.0,  5.5,  6.0,  6.5,  7.0,  8.0,  9.0,  10.0,  11.0,  12.0,  13.0,  14.0,  15.0,  16.0,  18.0,  20.0, 50.0, 99.0   \\ \hline 
$12^4$ &  4.5,  4.7,  5.0,  5.3,  5.5,  6.0,  6.5,  7.0,  8.0,  9.0,  10.0,  12.0,  14.0,  16.0,  18.0,  20.0, 50.0, 99.0   \\ \hline 
$16^4$ &  5.3,  5.5,  5.7,  6.0,  6.5,  7.0,  7.5,  8.0,  9.0,  10.0,  12.0,  14.0,  16.0,  18.0, 20.0, 50.0, 99.0   \\ \hline 
$20^4$ &  5.7,  6.0,  6.5,  7.0,  8.0,  9.0,  10.0,  12.0,  14.0,  16.0,  18.0,  20.0, 50.0   \\ \hline 
\end{tabular}
\caption{Summary of simulation parameters}
\label{table:numtraj}
\end{table}
\section{Results}\label{sec:results}
\begin{figure}[]
\hspace*{5mm}\\
(a)\hspace*{83mm}(b)\vspace*{-7mm}\\
\hspace*{2mm}
\scalebox{0.65}{\includegraphics{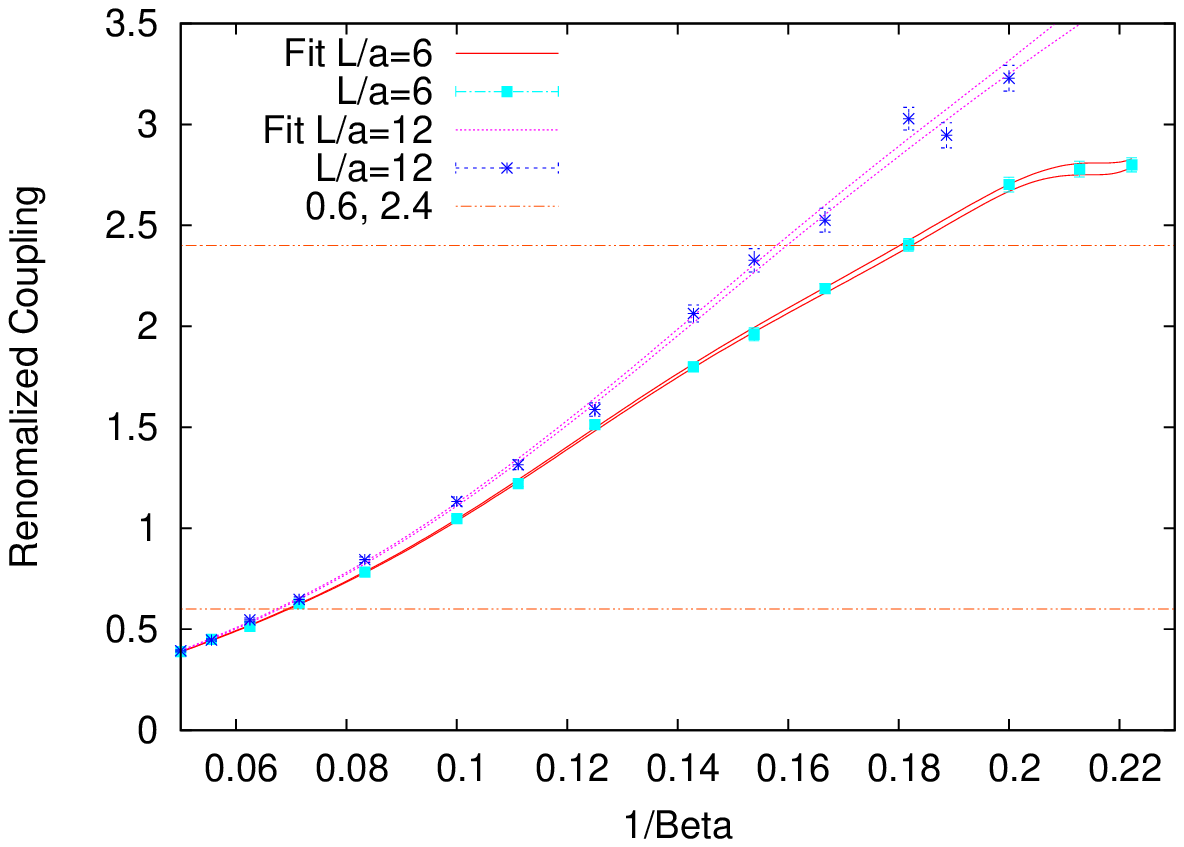}}
\hspace*{3mm}
\scalebox{0.65}{\includegraphics{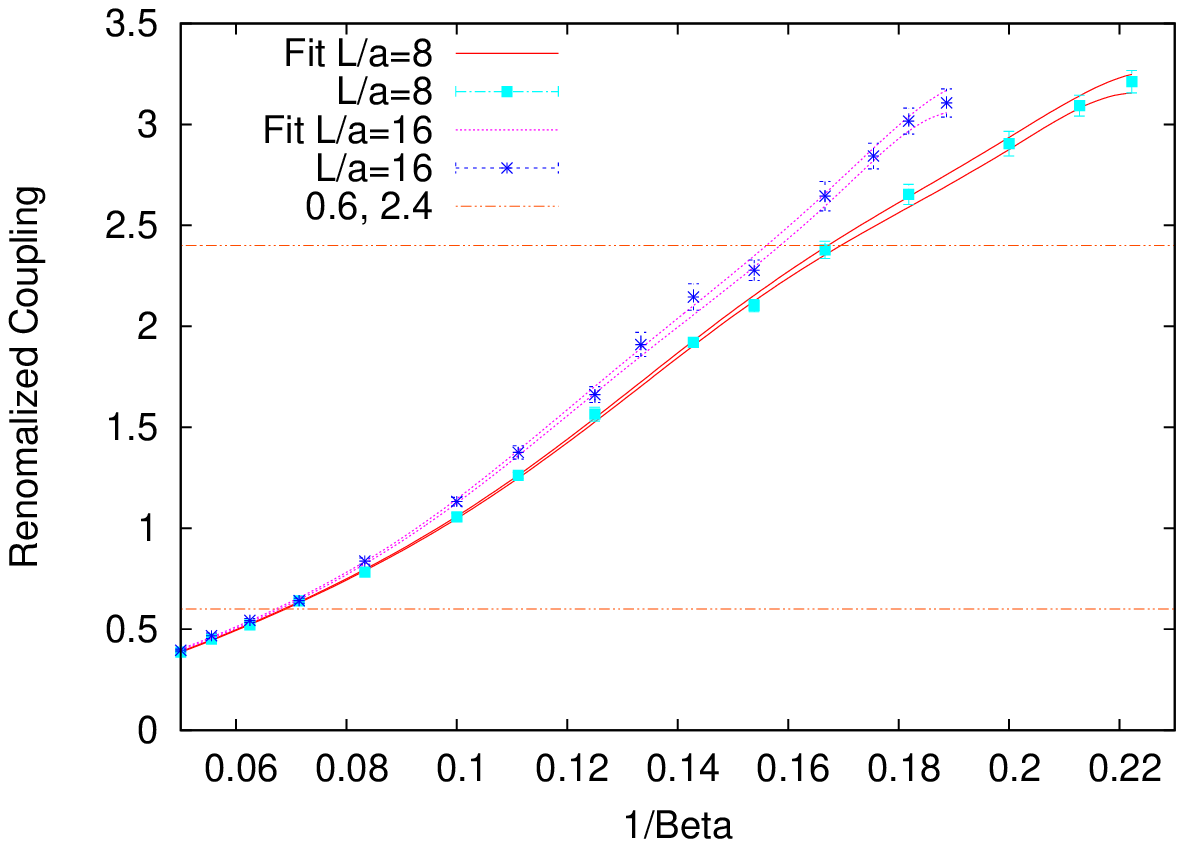}}
\vspace*{1mm}\\
(c)\hspace*{83mm}(d)\vspace*{-7mm}\\
\hspace*{2mm}
\scalebox{0.65}{\includegraphics{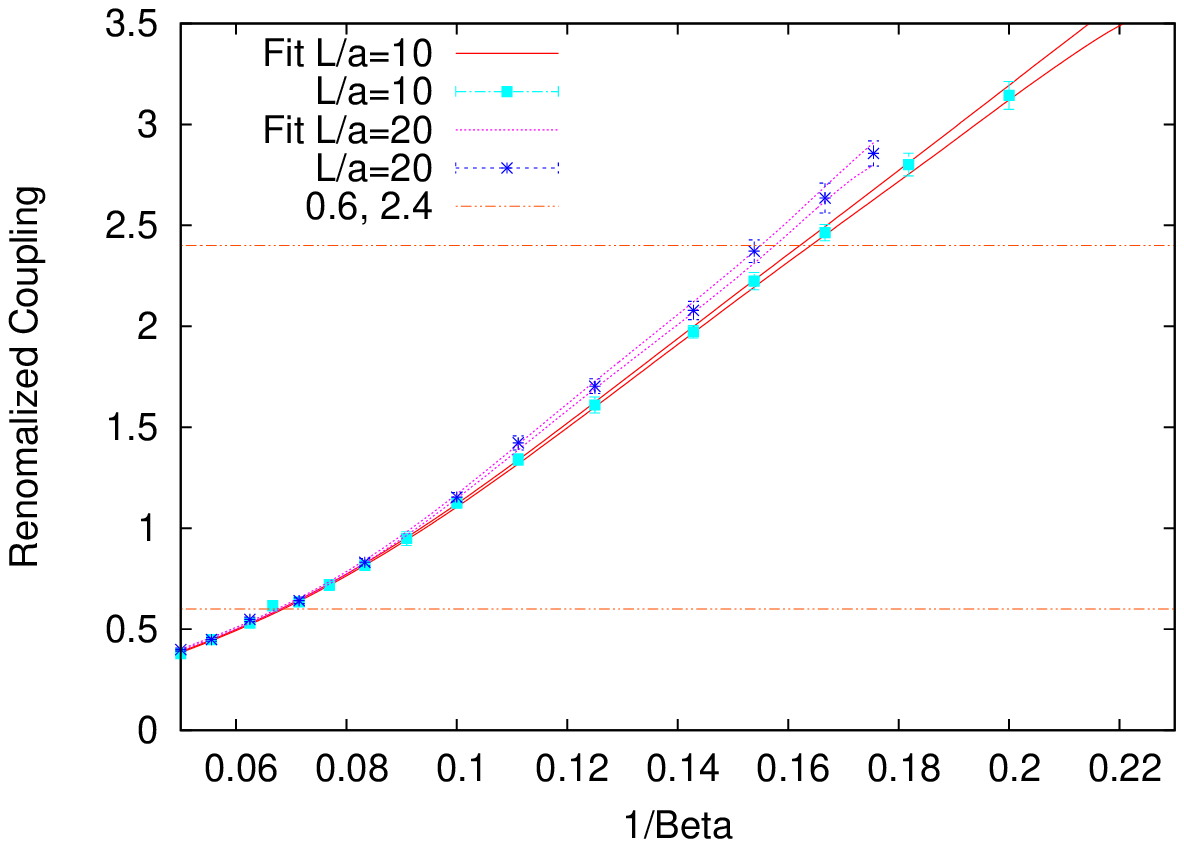}}
\hspace*{3mm}
\scalebox{0.65}{\includegraphics{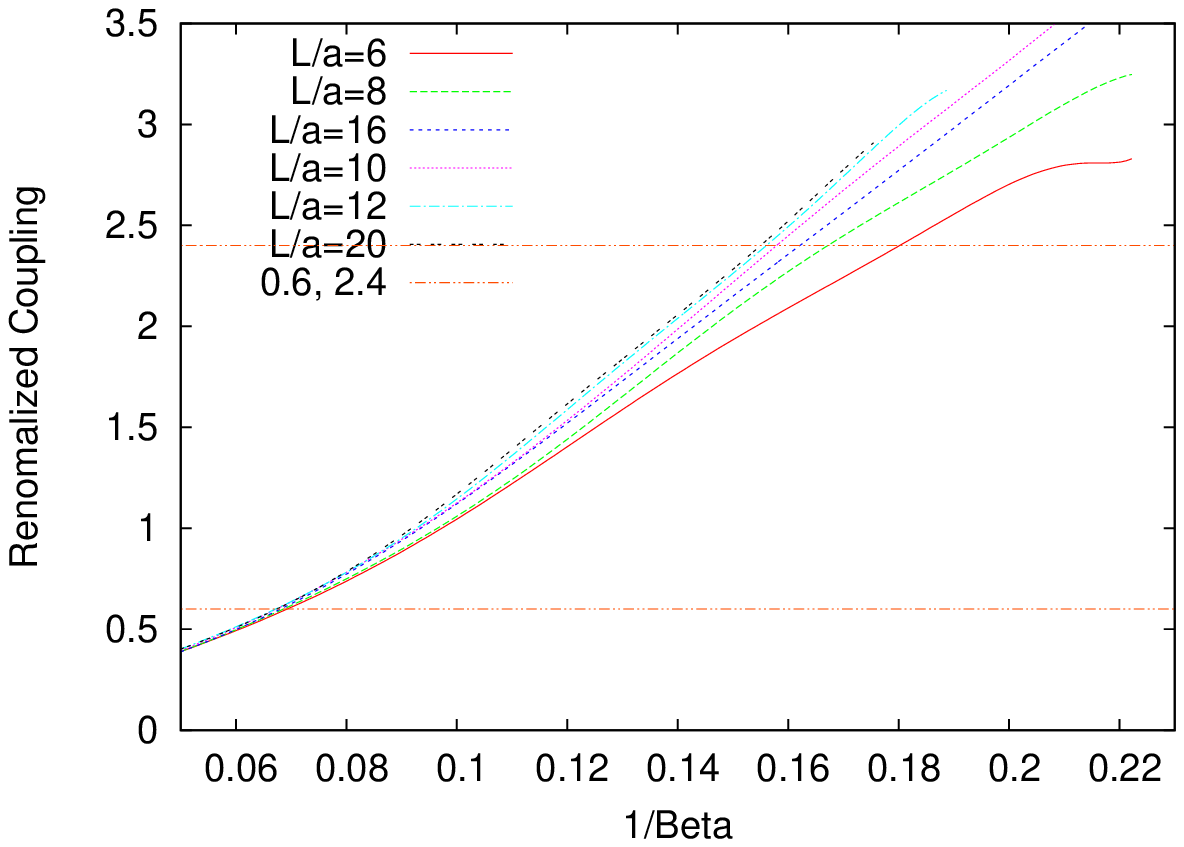}}
\caption{(a,\,b,\,c) - The renormalized coupling from the simulation are shown by points with error bars. 
Fit functions ($\pm$ error by $\Delta \chi^2 = 1$) are shown as curves. The lattice sizes $L/a$ are  $(6,12)$, $(8,16)$ and $(10,20)$, respectively.
The value of input coupling $u$ for the step scaling function at the left and right edges in Fig.\,2-c, 0.6 and 2.4, are shown for the guide of eyes. (d) - Central values of the fitting function for all sizes are shown.}
\label{fig:coupling}
\end{figure}

\begin{figure}[]
(a)\hspace*{83mm}(b)\vspace*{-7mm}\\
\hspace*{2mm}
\scalebox{0.65}{\includegraphics{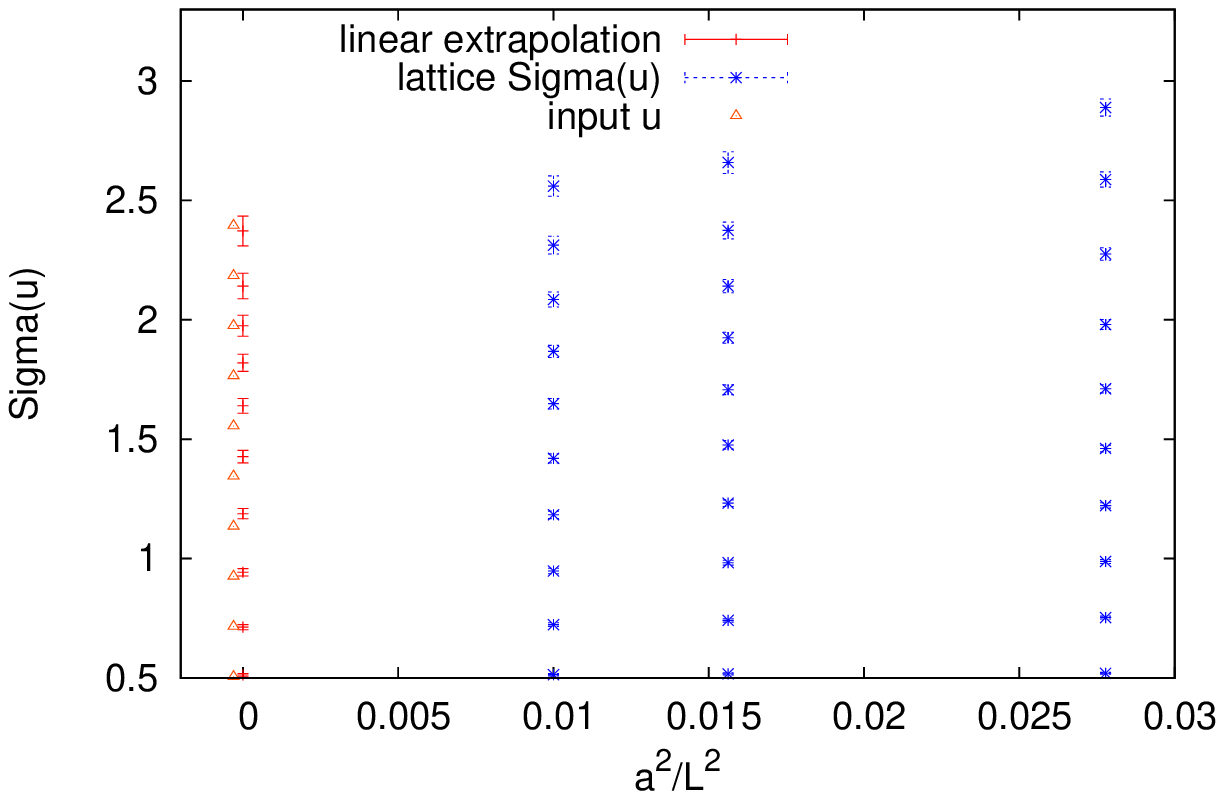}}
\hspace*{3mm}
\scalebox{0.65}{\includegraphics{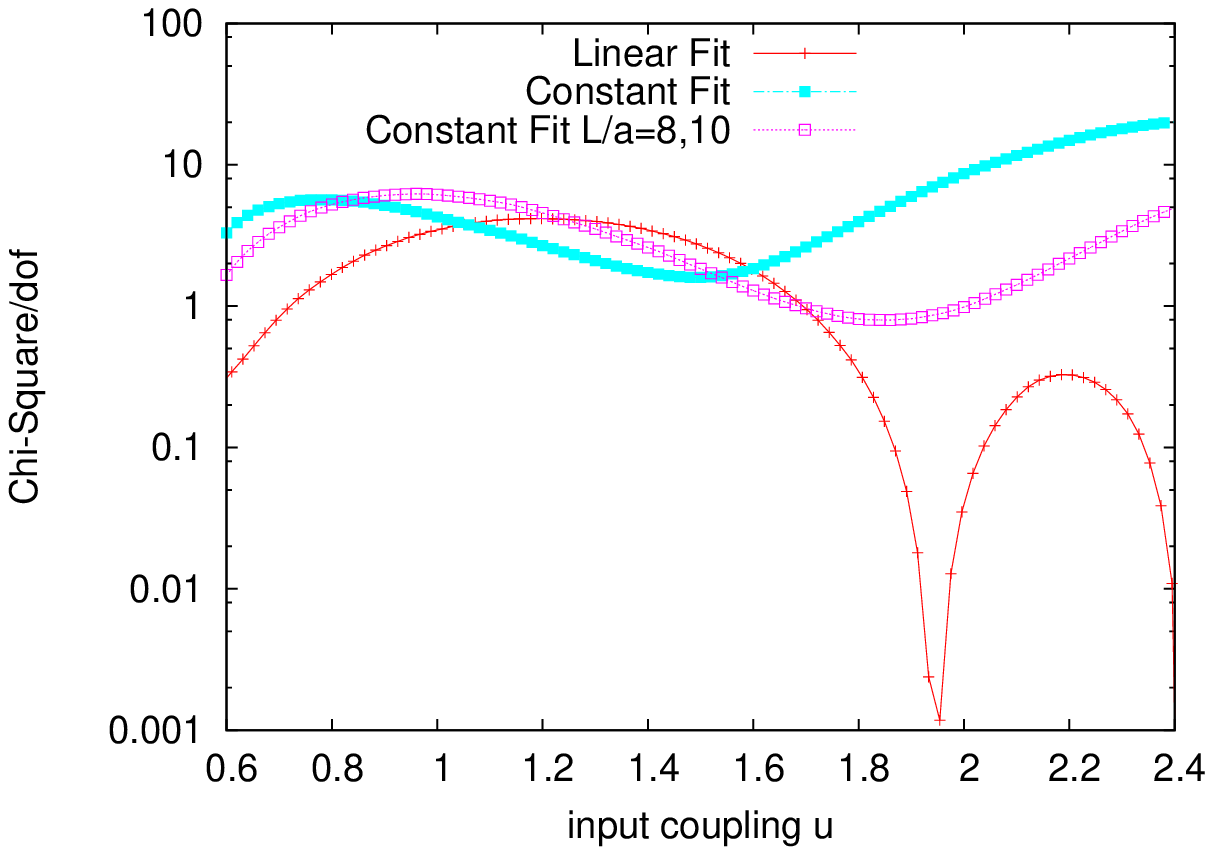}}
\vspace*{1mm}\\
\hspace*{41mm}(c)\vspace*{-7mm}\\
\hspace*{41mm}
\scalebox{0.65}{\includegraphics{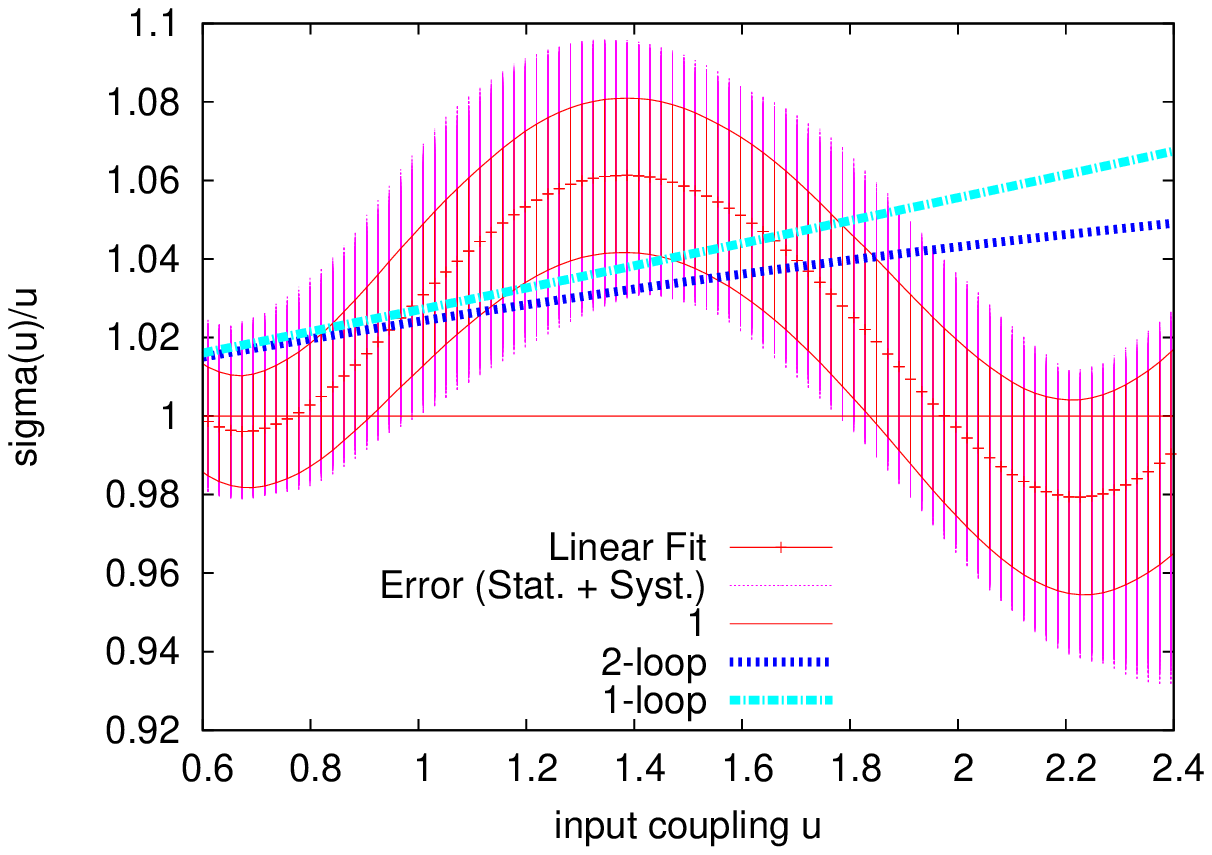}}
\caption{(a) - The lattice step scaling function $\Sigma( u , L/a )$ and the step scaling function in the continuum limit $\sigma(u)$ with linear fit. (b) - The quality of fit $\chi^2/(\mathrm{d.o.f})$ for linear fit, constant fit and constant fit with only finer lattice. (c) - Growth ratio $\sigma(u)/u$ with linear fit. Systematic errors are given by changing the degree of polynomial in fit function. The result of perturbative calculation using one-loop and two-loop are also shown. --- Step size is 2 for all panels (a),(b),(c).}
\label{fig:ssf}
\end{figure}

In Fig.\,\ref{fig:coupling},
we show the $\beta$-interpolation of renormalized coupling constant.
The degree of the polynomial $ ( 2 n + 1 ) $ in Eq.\,(\ref{eq:fitfunc}) is chosen, at fixed $L/a$, such that 
it leads to the smallest $\chi^2/\mathrm{(d.o.f.)}$, as shown in Table\,\ref{table:fitchisq}.
From Fig.\,\ref{fig:coupling}-a to Fig.\,\ref{fig:coupling}-c, it is clear that the smaller $L/a$ is, the larger the difference between 
$\bar g^2_\mathrm{latt}(\beta,2L/a)$ and $\bar g^2_\mathrm{latt}(\beta,L/a)$ is.
This is particularly obvious in the infrared regime $\bar g^2_\mathrm{latt} > 2$.
It means that, for a given input coupling $u$, the lattice step scaling function is smaller for the larger $L/a$, as shown in Fig.\,\ref{fig:ssf}-a.
From Fig.\,\ref{fig:coupling}-d, we see that $\bar g_\mathrm{lattice}^2( \beta , L'/a ) > \bar g_\mathrm{lattice}^2( \beta , L/a ) $ for
$L'/a > L/a$ except for ultraviolet region.
This means, outside the asymptotic-freedom regime, the lattice step scaling function $\Sigma(u,s,L/a)$ 
is larger than $u$ for any step size $s$.

In Fig.\,\ref{fig:ssf}, the lattice step scaling function and the continuum limit are shown.
From the Fig.\,\ref{fig:ssf}-b, we see that linear fit gives much smaller $\chi^2/({\rm d.o.f})$ in the region $u = 1.8 \sim 2.4$, 
on the other hand, constant fit gives a larger value of $\chi^2/({\rm d.o.f})$ $\sim$ 0.8 to 20 in that region.
For $u = 1 \sim 1.7$, the $\chi^2/({\rm d.o.f})$ from constant fit and linear fit are about the same.
From these, we conclude that the linear fit is much preferable than the constant fit.
In Fig.\,\ref{fig:ssf}-c, we show the growth ratio $\sigma(u)/u$.
To estimate the systematic error, we changed the number of fit parameters in the $\beta$ interpolation Eq.\,(\ref{eq:fitfunc}),
from the choices given in Table\,\ref{table:fitchisq} by $\pm 1$ for all lattice sizes. We calculated the step scaling function using all possible $3^6$ combinations for this interpolation.
The narrow (red) band in Fig.\,\ref{fig:ssf}-c is the statistical error associated
with the procedure using the best fit (with parameters in Table\,\ref{table:fitchisq}) for
the $\beta$ interpolation.  On the other hand, the wide (purple) band
indicates the largest deviation (including the statistical
fluctuations) from the central value of the best fit in all the
$\beta$ interpolations that we carried out.   Therefore, this band
contains both statistical and systematic errors.
In Fig.\,\ref{fig:ssf}-c, it shows that $\sigma(u)/u$ is consistent with one in $1.79 < u $.
This supports that this theory contains an infrared fixed point.
\begin{table}[]
\centering
\begin{tabular}{|c|c|c|c|c|c|c|} \hline
Size $L/a$ & 6 & 8 & 10 & 12 & 16 & 20 \\ \hline
Number of points & 17 & 17 & 19 & 18 & 17 & 13 \\ \hline
Number of fitting parameters & 8  & 8  & 6  & 5  & 8  & 7  \\ \hline
Degree of polynomial         & 19 & 19 & 15 & 13 & 19 & 17 \\ \hline
$\chi^2/\mathrm{d.o.f.} $ & 1.04 & 0.97  & 0.59 & 1.65 & 0.54 & 0.61 \\ \hline
\end{tabular}
\caption{$\chi^2/\mathrm{d.o.f.}$ for the fitting of beta.}
\label{table:fitchisq}
\end{table}
\section{Summary and Comments}\label{sec:summary}
In this work, we show the existence of the infrared fixed point\,(IRFP) in SU(3) gauge theory with 12 massless fermions in the fundamental representation,
by using the step scaling method with the Twisted Polyakov Loop scheme.
The results presented in this article are obtained with step size $s = 2$.
The location of IRFP reported here is consistent with our work \cite{Aoyama:2011ry} using step size $s=1.5$.

The lattice step scaling function exhibits $\Sigma( u , L/a ) > u$ at finite lattice spacing, except for the ultraviolet region.
On the other hand, the step scaling function in the continuum limit shows that $\sigma(u) \sim u$ in the infrared region.
This property at finite lattice spacing is different from the results in the step scaling function study using the Schroedinger Functional scheme\,\cite{Appelquist:2009ty}\cite{Appelquist:2007hu},
 in which $\Sigma( u , L/a ) \sim u$ in the infrared region even at the finite lattice spacing. 
%
\section*{acknowledgments}
We thank G. Fleming, Y. Taniguchi, and N. Yamada for useful discussions. E. I. would like to express her gratitude to H. Terao for the detailed discussions. 
The numerical simulations were carried out on NEC SX-8 and Hitachi SR16000 at YITP, Kyoto University, NEC SX-8R at RCNP, Osaka University, and Hitachi SR11000 and IBM System Blue Gene Solution at KEK under its Large-Scale Simulation Program (No. 09/10-22 and 10-16), as well as on the GPU cluster at Taiwanese National Center for High-performance Computing.
We acknowledge Japan Lattice Data Grid for data transfer and storage.
This work is supported in part by the Grant-in-Aid of the Ministry of Education (Nos. 20105002, 20105005, 21105508, 22740173, and 23105708 ), and the National Science Council of Taiwan via grants 99-2112-M-009-004- MY3 and 099-2811-M-009-029.

\end{document}